\documentstyle[prl,aps,epsf,epsfig,multicol]{revtex}

\newcommand{\be}{\begin{equation}}
\newcommand{\ee}{\end{equation}}
\newcommand{\bes}{\begin{eqnarray}}
\newcommand{\ees}{\end{eqnarray}}
\newcommand{\bma}{\left( \begin {array}}
\newcommand{\ema}{\end {array} \right)}

\begin{document}

\title{Synchronization of Coupled Systems with Spatiotemporal Chaos}

\author{Peter Grassberger}

\address{HLRZ c/o Forschungszentrum J\"ulich, D-52425 J\"ulich, Germany}

\date{\today}

\maketitle
\begin{abstract}  
We argue that the synchronization transition of stochastically coupled cellular 
automata, discovered recently by L.G. Morelli {\it et al.} (Phys. Rev. {\bf 
58 E}, R8 (1998)), is generically in the directed percolation universality 
class. In particular this holds numerically for the specific example studied by 
these authors, in contrast to their claim. For real-valued systems with 
spatiotemporal chaos such as coupled map lattices, we claim that the synchronization 
transition is generically in the universality class of the Kardar-Parisi-Zhang 
equation with a nonlinear growth limiting term.

  \vspace{4pt}
\end{abstract}

\begin{multicols}{2}

Since the pioneering work of Fujisaka {\it et al} \cite{fuji1,fuji-yama,yama,fuji-ishi} 
and others \cite{pikov,waller,schuster}, synchronization of chaotic systems has become 
a very intensely studied subject, partially due to hopes that this could lead to 
applications in control and secure communications \cite{carrol}.

In spatially extended systems, synchronization can appear in (at least) two 
forms. On the one hand, one can ask whether distant regions in a single such 
systems can oscillate in phase. After this phenomenon was observed by 
Chat\'e {\it et al} \cite{cm} in high dimensional cellular automata, it was 
realized that it can be mapped onto the Kardar-Parisi-Zhang (KPZ) problem of the 
growth of a random surface \cite{kpz}, with a synchronized state corresponding to 
a globally smooth phase $\phi = \phi({\bf x})$ \cite{gallas,mukamel,cgt}. 

In the present paper we shall deal with another problem, namely that of mutual 
synchronization of two identical locally coupled systems. If these systems are 
described by state variables $x_i^t$ and $y_i^t$ (for simplicity we assume here 
discrete 1-dimensional space $i$ and discrete time $t$), we write the evolution 
in general as 
\bes
   x_i^{t+1} = f(\ldots x_{i-1}^t,x_i^t,x_{i+1}^t\ldots) + \epsilon g(x_i^t-y_i^t), \\
              \nonumber
   y_i^{t+1} = f(\ldots y_{i-1}^t,y_i^t,y_{i+1}^t\ldots) + \epsilon g(y_i^t-x_i^t).
\ees
The function $f$ is nonlinear such that the evolution is chaotic for $\epsilon = 0$.
Due to sensitive dependence on initial conditions, ${\bf x}^t$ and ${\bf y}^t$ 
will be completely uncorrelated in this case, unless they started with identical 
initial conditions.  
Synchronization should only be expected for $\epsilon>0$ if $g(x)$ is negative 
for positive small $x$, so that any small difference $x_i^t-y_i^t$ will be damped 
by the last terms in eq.(1). 

For chaotic systems with a finite number of degrees of freedom, there is a finite 
synchronization threshold $\epsilon_c$, 
with intermittent behavior and `riddled' \cite{yorke} attractor basins near 
$\epsilon=\epsilon_c$ \cite{fuji1,fuji-yama,yama,fuji-ishi,pg}. 
Recently, it was found numerically that essentially the same phenomena occur in 
spatially extended systems \cite{morelli,zanette,amengual}. While chaos in 
systems with few degrees of freedom requires $x^t$ to be real-valued, spatio-temporal 
chaos can occur also in systems with discrete $x_i^t$, co-called cellular 
automata (CA).
Therefore one can ask whether the phenomenon of mutual synchronization can 
occur also in the latter \cite{Footnote}, and whether there are universal 
scaling laws at the synchronization threshold which apply both to real-valued 
systems (coupled map lattices and partial differential equations) 
and to CA. While the former question was asserted positively in \cite{morelli}, 
we shall argue that the latter has a negative answer. For CA, the synchronization
threshold is generically in the directed percolation universality class 
\cite{dp-grass}, while it is for continuous systems in the universality class 
of KPZ growth with a nonlinear growth limiting term 
\cite{kurths-piko,dp-grass,grinstein,munoz}.

Let us first study the case of 1-d cellular automata. The specific system studied 
in \cite{morelli} was two copies evolving according to Wolfram's \cite{wolfram} 
rule 18 with periodic boundary conditions, and endowed with an additional 
stochastic coupling term. In rule 18, 
$x_i^t$ can assume two values 0 or 1, and the evolution function $f$ depends 
only on $x_i^t$ itself and its two nearest neighbors, 
$f(0,0,1)=f(1,0,0)=1$ and $f(x_{i-1},x_i,x_{i+1})=0$ else. The coupling was 
realized as follows: after applying the above rule to both ${\bf x}$ and ${\bf y}$, 
it was checked whether $x_i=y_i$. If not, a random number is drawn uniformly 
from $[0,1]$. If this number is less than some fixed number $p$, a second 
random number is drawn and, depending on that, either  
$x_i$ is put equal to $y_i$ or $y_i$ is put equal to $x_i$. Thus $x_i=y_i$ is 
enforced with probability $p$, while both $x_i$ and  $y_i$ are left untouched with
probability $1-p$. It was found numerically that the system synchronizes for 
$p>p_c = 0.193$. For $p<p_c$ the density of sites with $x_i\neq y_i$ scales for 
$t\to\infty$ as $(p_c-p)^\beta$ with $\beta = 0.34$, while it decays for 
$p>p_c$ with a characteristic time $T$ which scales as $T\sim (p-p_c)^{-\nu_\|}$ 
with $\nu_\|=1$. Since these exponents disagree grossly with the DP values 
$\beta = 0.2765,\; \nu_\| = 1.7338$ \cite{jensen}, it was concluded 
that this transition is not in the DP universality class.

Unfortunately, the above estimates are flawed for several reasons. The first 
is that it is notoriously difficult to measure $\beta$ directly in DP and similar 
processes, due to the very slow convergence towards the stationary state. At 
$p=p_c$, the density of `active' sites in DP (corresponding to sites with 
$x_i\neq y_i$ in the present model) scales as $\rho\sim t^{-\delta}$ with $\delta 
= 0.1595$ \cite{jensen}. Near $p_c$, this means that one has to wait excessively 
long until the stationary state is reached. Much more reliable results are obtained 
by following the approach towards the asymptotic state, e.q. by measuring the 
decay of $\rho$ with time. 

The second problem is that rule 18 is well known to have very slow convergence 
towards its asymptotic state \cite{grass-r18}, in contrast to claims made in 
\cite{morelli}. Therefore the strategy of discarding a transient of a few 
hundred time steps used in \cite{morelli} is bound to induce errors.
When starting with a random initial state, rule 18 orders into domains in which 
$x_i^t$ is zero either for even $i+t$ or for odd $i+t$. The boundaries between 
these domains move according to annihilating random walks, so that the 
domain sizes grow $\sim \sqrt{t}$ and the density of domain walls decreases 
as $1/\sqrt{t}$. Asymptotically, the entire lattice is one single domain. On 
the sublattice where $x_i^t$ is not identically zero, its evolution follows the 
`additive' rule 90 given by $f(0,0,1)=f(1,0,0)=f(0,1,1)=f(1,1,0)=1$. The 
invariant state of the latter is completely random. Therefore rule 90 must 
show the same synchronization threshold as rule 18 and the same critical exponents, 
but it involves {\it no} transient whatsoever if one starts with random initial 
conditions.

\begin{figure}[b]
  \begin{center}
    \psfig{file=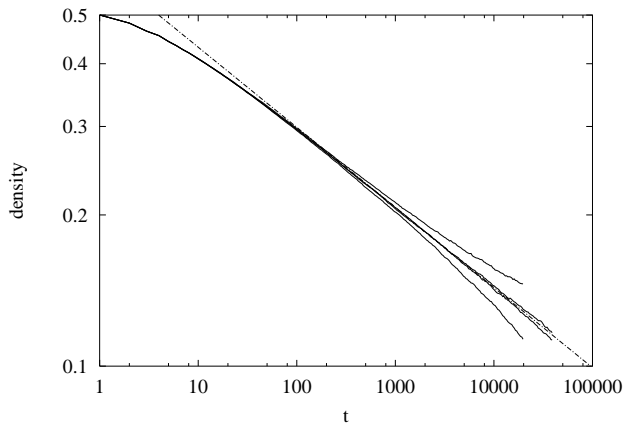,width=6.cm,angle=270}
    \begin{minipage}{8.5cm}
      \caption{Log-log plot of $\rho(t)$ for rule 90, for several values 
       of $p$: $0.1902, 0.19059,0.19065,0.1910$ from top to bottom. Statistical 
       errors are smaller than the thickness of the lines. The straight dashed 
       line has slope $-0.1595$ as predicted by DP.
        }
  \end{minipage}
\end{center}
\label{fig1}
\end{figure}

The density $\rho(t)$ for rule 90 is shown in Fig.1 for several values of $p$. 
To obtain these data we used ca. 1000 lattices of size $L=10000$ for each $p$, 
which gave a sample more than 100 times larger that that of \cite{morelli}.
We see clearly a power behavior for large $t$ (with strong small-$t$ corrections) 
for $p=p_c = 0.19061\pm 0.00003$. This is quite far from the value given in 
\cite{morelli} and implies immediately that the inserts in Figs. 2 and 3 of 
that paper are very misleading. We also see from Fig.1 that $\delta=0.159\pm 0.003$
in excellent agreement with DP. After having determined $p_c$ in this way, we 
performed very long runs (with up to 400,000 time steps and with $L$ up to 40,000)
for $p<p_c$ in order to estimate $\rho(t=\infty)$. Such long runs were needed 
since otherwise we would have suffered from systematic errors. Results are shown 
in Fig.2 and give $\beta = 0.277\pm 0.007$, again in perfect agreement with DP 
\cite{footnote2}. 

Finally, we show in Fig.3 the density decay $\rho(t)$ for rule 18, with 
random initial conditions and without discarding any transient. We see indeed 
rather large corrections to scaling for times $\gg 10^2$. If we would locate 
the critical point by means of least square fits including the short time region,
we would systematically overestimate $p_c$ and $\delta$, just as was done in 
\cite{morelli}.

\begin{figure}[b]
  \begin{center}
    \psfig{file=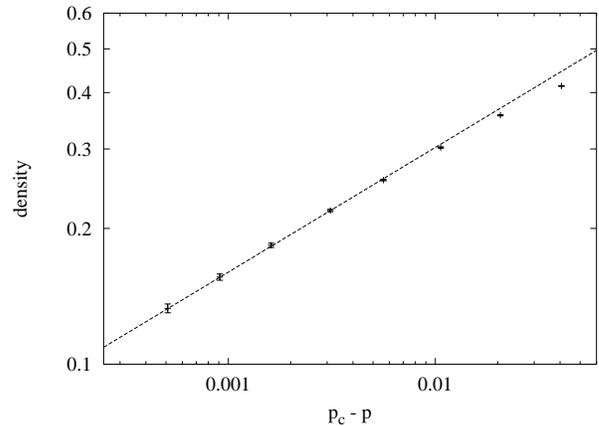,width=6.cm,angle=270}
    \begin{minipage}{8.5cm}
      \caption{Log-log plot of $\rho(t=\infty)$ against $p_c-p$, with $p_c$ 
       as obtained from Fig.1. The dashed line has slope $0.2765$ 
       as predicted by DP.
        }
  \end{minipage}
\end{center}
\label{fig2}
\end{figure}

\begin{figure}[b]
  \begin{center}
    \psfig{file=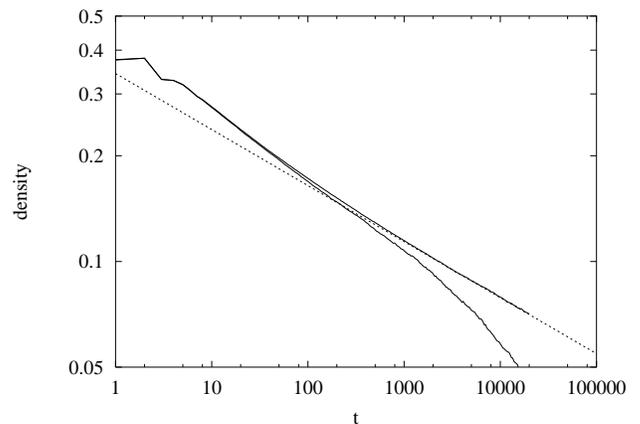,width=6.cm,angle=270}
    \begin{minipage}{8.5cm}
      \caption{Same as Fig.1, but for rule 18 and for two values of $p$ only: 
       0.1906 (top) and 0.192 (bottom). The first gives perfect agreement 
       with DP for large $t$ but large deviations at small $t$. The second 
       would give a better least square fit to a straight line for $10<t<5000$, 
       but this would yield wrong estimates of $p_c$ and $\delta$.
        }
  \end{minipage}
\end{center}
\label{fig3}
\end{figure}

We also studied rule 22. This is not an additive rule, and it has 
a nontrivial invariant measure with maybe zero entropy, but non-zero 
Lyapunov exponent \cite{rule22} (for the notions of chaos, entropy and Lyapunov 
exponents for CA see \cite{wolfram2}). Even if the entropy is not zero, there are very 
long ranged correlations in the invariant measure of rule 22 \cite{rule22}. 
It is thus of interest to see whether there is still a synchronization transition, 
and whether it is still in the DP universality class. 
We found again perfect agreement with DP. The 
critical point is at $p_c=0.22735\pm 0.00005$. It is easily seen that the sum of 
right- and left-moving Lyaponov exponents has to be positive for $p_c$ to be 
non-zero in any 1-dimensional CA. But the above values for $p_c$ and the 
known values for the Lyaponov exponents for rules 22 and 90 \cite{rule22,wolfram2} 
suggest that there is no simple relationship beyond this qualitative criterion.

On the theoretical side we also have no formal proof of the universality with DP, 
but we can use exactly the same heuristic arguments which were used in \cite{damage} 
to argue that damage spreading transitions are generically in the DP universality 
class. We refer to \cite{damage} for a detailed discussion, including caveats 
and limitations of the expected universality. 
Our present results underline again the remarkable robustness of DP critical 
behavior. In contrast to a statement made in \cite{morelli}, up to now DP universality
was verified in all tested cases (even if the original authors often found violations, 
such as in the present case), provided the criteria listed in \cite{damage} were 
met.

Let us now shortly discuss systems with continuous variables such as 
coupled map lattices. The main difference between these and CA is that 
synchronization is never perfect for finite time, even if $\epsilon>\epsilon_c$. 
Instead, the differences $|x_i^t-y_i^t|$ decrease exponentially with $t$ when 
the systems synchronize. But this means that close to threshold statistical 
or chaotic fluctuations can make the system de-synchronize again, at least locally.
Technically spoken, the system does not enter an absorbing state when it 
synchronizes locally, in contrast to the discrete case. This implies very 
different scaling exponents, as first noticed in \cite{kurths-piko} and verified 
in \cite{dp-grass,grinstein,munoz}. The generic stochastic partial differential 
equation with these features contains a diffusion term, a local nonlinear term, 
and a multiplicative noise term. The logarithm of the field appearing in this 
stochastic PDE satisfies the KPZ equation with an additional term which prevents 
the height variable from overcoming a barrier which we can conveniently place at 
$h=0$. The synchronization transition in this version corresponds to a transition 
from a surface pinned at $h\approx 0$ (desynchronized state), to a surface 
drifting towards $h=-\infty$ (synchronized state). 

We conjecture that the 
transition found in the neural network model of \cite{zanette} is in this 
universality class. It would be interesting to make detailed simulations of 
that model to verify this numerically. 

On the other hand we conjecture that the synchronization transition studied in 
\cite{amengual} is {\it not} an this universality class. In \cite{amengual} the 
coupling strength was called $\gamma$, and perfect synchronization occurred exactly 
at $\gamma=1$. As seem from eq.(1) of \cite{amengual}, the equivariance group of 
the coupled system changes at $\gamma=1$. For $\gamma\neq 1$ the system is invariant 
under phase transformations $A_{1,2} \to A_{1,2}e^{i\phi_{1,2}}$ and under the 
exchange $A_1 \leftrightarrow A_2$. For $\gamma=1$ one has the additional 
symmetry under phase rotations $A_1\pm A_2 \to (A_1\pm A_2)e^{i\phi_\pm}$. 
As stressed in \cite{grass-piko}, (de-)synchronization is essentially a phenomenon 
of spontaneous symmetry breakdown. As in other transitions with spontaneous
symmetry breaking, the universality class of such transitions should depend crucially 
on the type of symmetry broken, and should be particular sensitive to symmetry 
changes at the critical point.

\end{multicols}


\begin{references}

\bibitem{fuji1} H. Fujisaka, Prog. Theor. Phys. {\bf 70}, 1264 (1983).
\bibitem{fuji-yama} H. Fujisaka and T. Yamada, Prog. Theor. Phys. {\bf 69}, 32 (1983);
    Prog. Theor. Phys. {\bf 74}, 919 (1985); Prog. Theor. Phys. {\bf 75}, 1087 (1986).
\bibitem{yama} T. Yamada and H. Fujisaka, Prog. Theor. Phys. {\bf 70}, 1240 (1983); 
    Prog. Theor. Phys. {\bf 72}, 885 (1984); Phys. Lett. {\bf 124 A}, 8 (1987).
\bibitem{fuji-ishi} H. Fujisaka, H. Ishii, M. Inoue, and T. Yamada, 
    Prog. Theor. Phys. {\bf 76}, 1198 (1986).
\bibitem{pikov} A. Pikovsky, Z. Phys. {\bf B 55}, 149 (1984).
\bibitem{waller} I. Waller and R. Kapral, Phys. Lett. {\bf 105 A}, 163 (1984).
\bibitem{schuster} H.G. Schuster, S. Martin, and W. Martienssen, Phys. Rev. {\bf 
    A 33}, 3547 (1986).
\bibitem{carrol} L.M. Pecora and T.L. Carroll, Phys. Rev. Lett. {\bf 64}, 821 (1990);
    Phys. Rev. {\bf A 44}, 2374 (1991)
\bibitem{cm} H. Chat\'e and P. Manneville, Prog. Theor. Phys. {\bf 87}, 1 (1992).
\bibitem{kpz} M. Kardar, G. Parisi, and Y.-C. Zhang, Phys. Rev. Lett. {\bf 56}, 
    889 (1986)
\bibitem{gallas} J.A.C. Gallas, P. Grassberger, H.J. Herrmann und P. Ueberholz, 
    Physica {\bf A180}, 19 (1992).
\bibitem{mukamel} G. Grinstein, D. Mukamel, R. Seidin, and C.H. Bennett, 
    Phys. Rev. Lett. {\bf 70}, 3607 (1993)
\bibitem{cgt} H. Chat\'e, G. Grinstein, and L.-H. Tang, Phys. Rev. Lett. {\bf 74}, 
    912 (1995)
\bibitem{yorke} J.C. Alexander, J.A. Yorke, Z. You, and I. Kan, Int. J. Bifurc. 
    \& Chaos {\bf 2}, 795 (1992).
\bibitem{Footnote} For CA we will in general need stochastic functions $f$ or $g$ to 
    obtain non-trivial and unique synchronization thresholds. 
\bibitem{pg} A. Pikovsky and P. Grassberger, J. Phys. {\bf A 24}, 4587 (1991).
\bibitem{morelli} L.G. Morelli and D.H. Zanette, Phys. Rev. {\bf E 58}, R8 (1998).
\bibitem{zanette} D.H. Zanette and A.S. Mikhailov, Phys. Rev. {\bf E 58}, 872 (1998).
\bibitem{amengual} A. Amengual, E. Hern\'andez-Garc\'ia, R. Montagne, and M. San 
    Miguel, Phys. Rev. Lett. {\bf 78}, 4379 (1997)
\bibitem{dp-grass} P. Grassberger, "Directed Percolation: Results and Open Problems"
     in 'Nonlinearities in Complex Systems',
     Proceedings of 1995 Shimla Conference on Complex Systems,
     eds. S. Puri {\it et al}. (Narosa Publishing, New Delhi, 1997).
\bibitem{kurths-piko} A.S. Pikovsky and J. Kurths, Phys. Rev. {\bf E 49},
   898 (1994)
\bibitem{grinstein} G. Grinstein, M.A. Mu\~noz, and Y. Tu, Phys. Rev. Lett. {\bf 76}, 
    4376 (1996)
\bibitem{munoz} Y. Tu, G. Grinstein, and Mu\~noz, Phys. Rev. Lett. {\bf 78}, 274 
    (1997)
\bibitem{wolfram} S. Wolfram, Rev. Mod. Phys. {\bf 55}, 601 (1983).
\bibitem{jensen} I. Jensen, J. Phys. {\bf A 29}, 7013 (1996)
\bibitem{grass-r18} P. Grassberger, Phys. Rev. {\bf A 28}, 3666 (1983).
\bibitem{footnote2} If we would not use the precise value of $p_c$ obtained from 
    the data shown in Fig.1, the data of Fig.2 alone would give much less precise 
    estimates, $p_c = 0.19057\pm 0.00008$ and $\beta = 0.267\pm 0.015$.
\bibitem{rule22} P. Grassberger, J. Stat. Phys. {\bf 45}, 27 (1986).
\bibitem{wolfram2} S. Wolfram, Physica {\bf D 10}, 1 (1984).
\bibitem{damage} P. Grassberger, J. Stat. Phys. {\bf 79}, 13 (1995).
\bibitem{grass-piko} P. Grassberger and A. Pikovsky, 
    "Symmetry Breaking Bifurcations in Chaotic Systems", in `Renormalization Group '91', 
    eds. D.V. Sirkov {\it et al.} (World Scientific 1992).

\end{references}
\end{document}